# Intrusion Detection Framework for SQL Injection

### Israr Ali
*Faculty of Engineering Sciences & Technology*
*Iqra University Main Campus*
Karachi, Pakistan
Israr.ali @iqra.edu.pk

### Syed Hasan Adil
*Faculty of Engineering Sciences & Technology*
*Iqra University Main Campus*
Karachi, Pakistan
hasan.adil @iqra.edu.pk

### Mansoor Ebrahim
*Faculty of Engineering Sciences & Technology*
*Iqra University Main Campus*
Karachi, Pakistan
mebrahim@iqra.edu.pk

*Abstract*—In this era of internet, E-Business and e-commerce applications are using Databases as their integral part. These Databases irrespective of the technology used are vulnerable to SQL injection attacks. These Attacks are considered very dangerous as well as very easy to use for attackers and intruders. In this paper, we are proposing a new approach to detect intrusion from attackers by using SQL injection. The main idea of our proposed solution is to create trusted user profiles fetched from the Queries submitted by authorized users by using association rules. After that we will use a hybrid (anomaly + misuse) detection model which will depend on data mining techniques to detect queries that deviates from our normal behavior profile. The normal behavior profile will be created in XML format. In this way we can minimize false positive alarms.

*Index Terms*—association rules; Intrusion detection anomaly detection; SQL Injection; Databases

## I. INTRODUCTION

Database-driven web applications have gotten to be broadly sent on the Internet. Companies use them to give an expansive scope of administrations to their clients. These applications with their databases regularly contain secret, or even confidential data, for example, a client and budgetary records. With the passage of time the accessibility of these applications has expanded, there has been a relating increment in the number and complexity of intrusions that target them. A standout amongst the most genuine sorts of intrusion against web applications is Structured Query Language (SQL) Injection attacks. These attacks are among the highest of the main vulnerabilities that a web application can have from intruders. As the name intimates, this kind of assault is controlled and targeted towards the database layer of the web applications. Most web applications are ordinarily developed in a two- or three-layered construction modeling as described in Figure 1.

SQL Injection is a sort of code-invasion assault in which an assailant uses uniquely created inputs to trap the database into executing aggressor detailed database orders. It can give the aggressors immediate access to the underlying databases of a web application, with that, the ability to leak, change, or even erase data that is put away on them. SQL Injections happen when information gave by a client is not appropriately approved and is incorporated specifically in a SQL query.

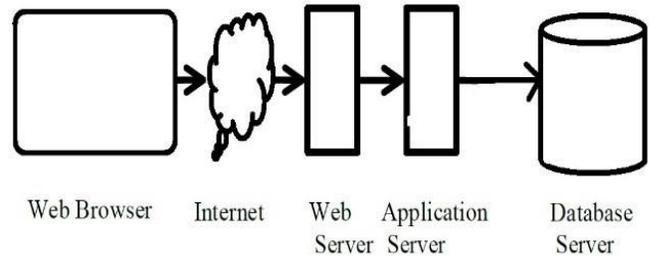

Fig. 1. Three Layered Architecture

We will give a basic case of Injection to show the issue. The attacker first tries to find the root path and Writable Directory on the website. For example, he will write in the address bar

**www.site.com/index.php?id=10'**

This will let him know whether the website is vulnerable to SQL Injection or not. Then to find the columns something like this can be used

**www.site.com/index.php?id=-10 Union Select 1, 2, 3, 4, 5—**

He can find the venerable column by

**www.site.com/index.php?id=-10 Union Select 1, 2, version (), 4, 5—**

Where we have 5 columns and columns three is vulnerable. After that the attacker can load a file or Shell by using

**www.site.com/index.php?id=-10 Union Select 1, 2, load file('/etc/my.cnf'),4,5—**

Then he can check the file privileges of the current user for this first he will find current username.

**www.site.com/index.php?id=-10 Union Select 1, 2, current user, 4, 5—**

This will return a user name that will be used in this statement

**www.site.com/index.php?id=-10 Union Select 1, 2, file priv, 4, 5 FROM mysql.user WHERE user=username-**

One component to protect against web assaults is to utilize intrusion detection systems and particularly network intrusion detection systems (NIDS). NIDS can use misuse detection (MD) or anomaly detection (AD) or both methods to safeguard against assaults. An intrusion detection systems (IDS) that utilize the anomaly detection (AD) method make a standard of typical usage examples, and anything that broadly strays from it gets hailed as a possible interruption. Misuse detection (MD) identification system utilizes particularly known examples of unapproved conduct to foresee and catch resulting comparable attack. These known examples are sometimes referred as Signature.

Tragically, NIDS are not productive or even helpful in web interruption detection. Since numerous web assaults concentrate on applications that have no confirmation on the underlying network or framework exercises, they are seen as an ordinary movement to the general NIDS and pass through them effectively. NIDS are basically sitting on the lower (network/transport) level of network model while web administrations are running on the higher (application) level. So, In this paper, we propose a new mechanism that joins together the two IDS procedures, AD and MD, to shield against SQL Injection Attacks. The primary thought of our SQL Injection Shield Framework (SSF) structure is to make a profile for web applications that can present to the normal behavior of users regarding SQL queries they submit to the database. Database logs could be utilized to gather these true blue questions given that these logs are free of interruptions. We then utilize SSF framework focused around data mining techniques to distinguish the queries that deviates from the profile of normal queries. The queries recovered from database log are put away in XML document with predefined structure. We pick XML Format on the grounds that it is more organized than level records, more adaptable than matrices, easier and devour less capacity than databases.

We can then use association rules on the XML File. These rules are description of the profile of typical conduct and any deviation from this profile will be considered an attack. Keeping in mind the end goal is to better distinguish SQL Injection Attacks and to minimize false positive alerts, SSF system as a second step uses misuse procedure to catch any change in the structure of the query. Vindictive clients now and then don't change the determination provision, however include an alternate SQL articulation or add particular essential words to the introductory query to check the helplessness of the site to SQL Injection Attacks or to perform inference attack. Such sorts of attacks are identified in the second step of the SSF. By looking at the structure of the query under test with the comparing queries in the XML document the past pernicious activities will be recognized.

Whatever remains of the paper will be sorted out as follows: in section II we will examine past work, Section III will give a point by point depiction about the SSF and its segments. AD and MD algorithms and a working case will be displayed in Section IV. Section IV finishes up the paper and diagrams future work.

## II. LITERATURE REVIEW

Diverse investigates and methodologies have been displayed to address the issue of web assaults against databases. Considering SQL Injection as top most risky assaults, as expressed in area I, there has been extraordinary research in identification and anticipation systems against this assault [1, 2, 3]. We can characterize these methodologies into two general classes:
a) One methodology is attempting to identify SQL Injection through checking peculiar SQL query structure.
b) An alternate methodology utilizes information conditions among information things which are more averse to change for recognizing noxious database exercises.

In both of two classifications, distinctive analysts exploit the profit of coordinating information mining with database interruption location keeping in mind the end goal to minimize false positive alarms, minimizing human intercession and better distinguish assaults [4]. In addition, Different interruption identification strategies are utilized either independently or together. Distinctive work utilized abuse procedure others utilized abnormality or blends the two systems.

Under the first class and without utilizing information mining method, Lee et al. in [5] and Low et al. in [6] created a structure focused around fingerprinting transactions for recognizing pernicious transactions. They investigated the different issues that emerge in the examination, representation and synopsis of this possibly enormous set of authentic transaction fingerprints. An alternate work that applies peculiarity identification strategy to distinguish odd database application conduct is exhibited by Valeur et al. in [7]. It constructs various distinctive factual question models utilizing a set of regular application questions, and after that captures the new queries submitted to the database to check for atypical conduct.

A general skeleton for recognizing noxious database transaction examples utilizing information mining was proposed by Bertino et al. in [8] [9] to mine database logs to structure client profiles that can display ordinary practices and recognize abnormal transactions in databases with part based access control instruments. The framework has the capacity distinguish interlopers by distinguishing practices that contrast from the typical conduct of a part in a database. Kamra et al. in [10] represented an upgraded model that can likewise distinguish gatecrashers in databases where there are no parts connected with every client. It utilizes bunching systems to structure succinct profiles speaking to typical client practices for distinguishing suspicious database exercises. An alternate approach that checks for the structure of the question to recognize malevolent database conduct is the work of Bertino et al. in [11]. They proposed a system focused around inconsistency recognition strategy and affiliation tenet mining to distinguish the query that goes astray from ordinary database application conduct.

The issue with this schema is that it delivers a considerable measure of guidelines and speaks to the questions in extremely enormous networks, which may influence hugely on the execution of standard extraction. Abuse discovery procedure have been utilized by Bandhakavi et al. in [12] to locate SQL Injection Attacks by finding the purpose of a question powerfully and afterward looking at the structure of the distinguished query with typical queries focused around the client data with the found plan. The issue with this methodology is that it needs to get to the source code of the application and make a few changes to the java virtual machine.

Halfond et al. in [13] created a strategy that utilize an approach to place illegal questions on the database. In its static part, the system uses program examination to regularly collect a model of the true request that could be made by the application. In its dynamic part, the method uses runtime seeing to survey the alterably created queries and check them against the statically-manufactured model. The framework WASP proposed by Wiliam et al. in [14] tries to counteract SQL Injection Attacks by a system called positive spoiling. In positive spoiling, the trusted piece of the question (static string) is not considered for execution and conceal as polluted, while all different inputs are considered. The trouble for this situation is the engendering of corrupts in an query crosswise over capacity calls particularly for the client characterized capacities which call some other outside capacities prompting the execution of a spoiled question. Distinctive different examines took after the same approach in location of bizarre SQL question structure in [15] [16].

Scrutinizes that fit in with the second classification of discovery which relies on upon information conditions are [17] [18] [19] [20]. The work that is focused around mining successive information access designs for database interruption discovery was proposed by Hu et al. in [17] [18]. Transactions that don't consent to govern created from read and compose arrangement sets are recognized as noxious transactions. Srivastava et al. offered a weighted question digging methodology [17] for catching database assaults. The playing point of the work displayed by Yihu et al. in [18] is the programmed disclosure and utilization of vital information conditions, to be specific, multi-dimensional and multi-level information conditions, for recognizing bizarre database transactions.

The commitment of this paper is a framework (SSF) that consolidates AD and MD procedure keeping in mind the end goal to better distinguish SQL Injection Attacks. This schema utilizes association standard rules with an AD system to manufacture the ordinary conduct of use, clients and locating irregular queries. Additionally, MD is utilized to check the structure of the query to recognize any noxious activities that can't be recognized utilizing AD method.

## III. THEORETICAL FRAMEWORK

Our Proposed framework (SSF) will detect intrusion before query execution at database end. For this reason we recommend to run this framework at the database end when all possible SQL injection attacks are meant to be detected. In this framework we will define a new hybrid approach of anomaly detection and misuse detection. The key thought of our structure is that we fabricate a store containing set of genuine questions submitted from the application client to the database. This store is actually our training dataset. We then utilize an inconsistency discovery methodology focused around data mining procedure to assemble a profile of ordinary application conduct and show queries that goes amiss from this typical conduct. In a second venture in the proposed framework we check for the presence of hazardous magic words in the query if the last breezes through the test of abnormality identification step. We require this step in light of the fact that in some cases the plan of the attacker is to distinguish the security gaps in the site or to derive the structure of the database through the slip message came back from the application. This sort of assault can't be caught through AD system in light of the fact that it doesn't oblige change in the states of the first query yet it will be found if the structure of the question is thought about against its comparing query in the storehouse document. Taking into account what a while ago expressed, we now know that the framework (SSF) works in two stages: one is the training stage and the other is detection stage. The accompanying subsections we will give a nitty gritty clarification of the schema, its parts and how it functions.

### A. Training Phase

In the training stage the training records are gathered from the queries the application send to the database. The source for getting these query traces is the database log gave that the latter is free of intrusions. The training stage stream is outlined in Fig. 3. The test here is that to proficiently encode these queries, keeping in mind the end goal to concentrate helpful peculiarities from them and appropriately construct the application finger impression. Not at all like methodology gave in [11], we decide to encode the questions in XML record. The encoding plan gave by Bertino et al. in [11] bring about a substantial, thick, scanty frameworks which may impact on the mining calculation. XML is more organized than flat records, is upheld by question tools like Xquery and XPath to concentrate information [21]. It is less difficult and expends less space than relational databases and more adaptable than grids.

It is vital to recognize precisely the structure of the XML record that will speak to the peculiarities separated from the queries that will help in building the application finger impression. The principle playing point of XML usage here is that XML nodes tags may be copied or duplicated upon need. For instance the number of ID Tags may contrast from one "query" Tag to an alternate relying upon the query itself. This is the reason it is more suitable to store questions than databases while keeping up adaptability and effortlessness. The XML document represented in the diagram, where this phase is depicted, actually stores the projection attributes and from clause of SQL Query and the predicate part of the query in a more itemized way. It is not vital to recognize the estimation of the whole integer number or string element but it is essential to establish that there is a number or string element or there is an alternate property in the right hand side of the SQL Query that is an effort of injection. An alternate record that ought to be made amid the preparing stage is the signature file that will be utilized amid the misuse location stage. As expressed before this record contains suspicious decisive words that may be viewed as an indication of intrusion detection by using SQL injection.

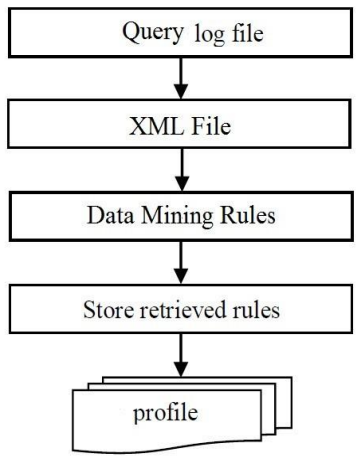

Fig. 2. Training Phase

Words like for instance single quote, semicolon, twofold dash, union, executive, request by and their hexadecimal representation with a specific end goal to keep the distinctive avoidance strategies [22] are signs of injections. The vital and the most important venture in the preparation stage is to fabricate the profile speaking to the application typical or normal behavior. The baseline from which we can decide what is good and what is wrong. We will apply association principles [23] on the XML file created from the database log record to concentrate and decides what speak to the ordinary or normal conduct of use. We found that different methodologies and approaches have been proposed to apply association rules on XML Nodes and data. We found that the approach presented by [24-26] for an in-depth overview of these methodologies is very useful and easy to use. The guidelines concentrated speak to relationship between each one table in the query with each one predicate in the determination condition.

This is focused around a perception that the static piece of the query is the projection characteristic and the part that is built amid execution is the determination part [11]. Here we add an alternate thing to the static part which are the tables in the from statement. After that we will attempt to make connections between the static part and the element part and concentrate guideline with backing and certainty of such connection. Any query that won't match tenets concentrated and put away with the standard profile will be considered assault. More insights about how the standards are concentrated are given in the accompanying subsection.

```
<AllQueries>
<Query no="1">
<commandType> select command </commandType>
<query_column> SSN </query_column>
<query_column> last_name </query_column>
<FromClause> Students </FromClause>
<LHS> first_name </LHS>
<RHS> string Literal </RHS>
<logical_operator> and </logical_operator>
<LHS> GPA </LHS>
<RHS> Integer Literal </RHS>
</Query>
</AllQueries>
```

Fig. 3. XML file containing Queries

## B. ANOMALY DETECTION PHASE

In the past subsection, we showed how the generous queries are gathered and caught in XML format in a structure empowering the framework we have proposed from making the database behavior profile. After this we can apply rules on the XML record containing real questions and concentrate decides that can depict the ordinary conduct of user that is they are normal users or attackers. The thought behind building the profile guideline is to apply one of affiliation rules calculations on a while ago made XML record to concentrate connection between each one table in the query with every determination characteristic barring the literals. In this way the standards concentrated have the accompanying format:

$$From \rightarrow LHS$$
$$From \rightarrow RHS$$

The rule that surpasses the base backing and certainty will be put away in a separate rule profile. These rules speak to the profile of how the application carries on typically. Fig. 5 shows the stream of recognition period of the schema when all is said in done including the anomaly method. When a database application is built, the fact and figures are usually supplied by the client develop the where clause of the query. In the meantime, the projection clause and the from clause stay static at the run time. So we make a connection between the static and the dynamic piece of the query and any change in the where clause by aggressors that can't be gotten from the standards profile will be published as an attack. We chose to pick the tables in the from clause from the static piece of the question rather than the projection characteristics on the grounds that the previous is more general and contain the most recent and therefore creating less controls and make it less demanding in correlation. Lets have another example of SQL injection attack

**Select username, password from admin where fname= or 1=1 - -**

Before executing this query, rules ought to be concentrated first furthermore contrasted with the standards in the normal user profile. The connection in the middle of tables and attributes will be analyzed against normal user profile put away in the profile guidelines record. The two relations under test from the past sample are:

Admin → username
Admin → 1

The primary connection exists in the normal user profile yet no such run the show matches the second one, so the query is published as SQL injection attack.

*C. MISUSE DETECTION PHASE*

In a second venture in the SQL injection detection process and after the anomaly recognition stage, comes the part of misuse identification. The need for this step originates from the way that SQL injection techniques doesn't just change the conditions in the query yet it additionally may give data about the database pattern or check the defenselessness of the application to SQL injection. This is carried out through adding to the query a few watchwords that may change the conduct of the question or return data about the database through database lapses without changing the predicates of the query. In such case, the anomaly identification stage won't have the capacity to find such assault. For example consider the following scenario

**Select * from admin where ID=10**

On the off chance that the aggressor simply includes a single quote toward the end of the question, this will bring about blunder message that may illuminate the assailant that the site is helpless against SQL Injection and he can perform his tricks. An alternate case of assault is simply including the essential word "order by" to the query without changing the determination characteristics like:

**Select * from admin where ID=10 order by 1**

Attempting to execute this query a few times will give attacker data about the quantity of fields in the table. This is why this step is required in the identification process. Additionally, our framework doesn't declare the query as abnormal just by discovering these keywords in the query on the grounds that it might be part of the genuine query itself bringing about false positive caution. This is the reason the proposed framework checks for the structure of the query under test with the comparing query put away in XML record. The recognition stage stream of the framework in Fig. 5 represents this methodology. These suspicious magic words are put away in record called magic words". This record contains SQL decisive words like single quote, request by, union select, semicolon, executive and their hexadecimal representation to maintain a strategic distance from the diverse avoidance systems. After affirming the presence of one or a greater amount of these pivotal words, we use XQuery to recover questions from XML file with the same projection attributes and same from clause. At that point examination is carried out between query under test and the query recovered by Xquery from XML record. In the event that there is no match then the query is advertised abnormality.

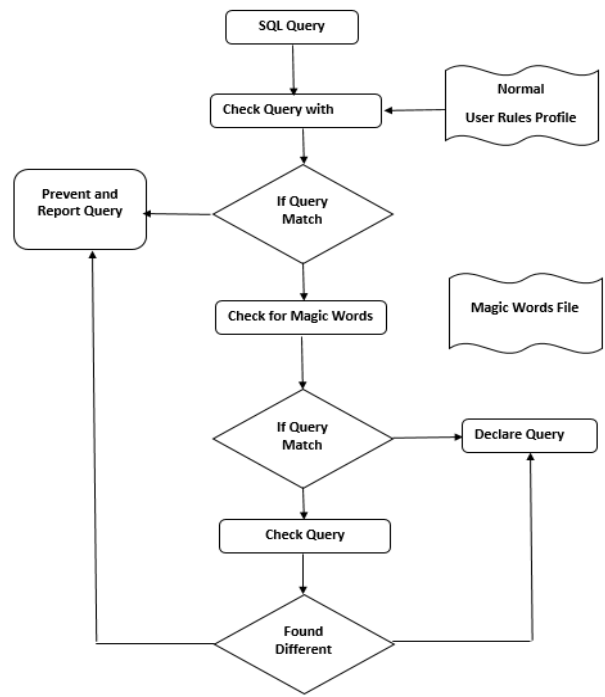

Fig. 4. Proposed framework flow

## IV. ALGORITHM AND WORKING EXAMPLE

In this segment we exhibit calculations for anomaly as well as misuse recognition. What's more, we give a working illustration outlining how the proposed framework skeleton performs the detection.

*A. Algorithm for anomaly detection*

**Inputs:**
1. Rules from Normal profile
2. Query submitted

**Output:**
1. True if query is intrusion
2. False if query is not intrusion

**Start:**
1. Fetch relationship between tables and selection fields from Query
2. Save Fetched relations in query relation array
3. Iterate each relation r in query relation array
    a. If (r is found in normal user profile(r) )
       i. score=score+1
    ii. If score = =length of query relation array
       Return false
    iii. Else
       Return true
4. End

*B. Misuse detection algorithm*

**Inputs:**
1. Magic keywords file
2. Query under test
3. XML file

**Output:**
1. True if query is intrusion
2. False if query is not intrusion

**Start:**
1. Iterate each keywords m in Magic keywords File
2. If k not exists in Query
    a. Return false
3. Else
    a. Use XQuery language to extract relevant queries from XML file
    b. If query structure doesnt match any retrieved queries
       Return True
    c. Else
       Return false
4. End

## V. WORKING EXAMPLE

To give better understanding of the anomaly and misuse identification in proposed framework system, we give in this subsection illustration of the stream of intrusion detection either inconsistency or misuse in this framework. The accompanying speaks to illustration of queries submitted from application to database:

- Select username, password from admin where id=?
- Select username, password from admin where id<?
- Select * from admin where username=? order by username
- Select username, product from admin where salary<? and IsActive=?

Our proposed framework will generate XML file like this

Fig. 5. XML file representing queries

In the wake of applying affiliation rules calculation like for instance Apriori on this XML record, the ensuing principles will put away in standards profile document like in Fig. 6. In the accompanying we will give specimen of vindictive and malicious queries.

- Select username, password from admin where id=5

The initial phase in the system is to distinguish connection in the middle of tables and selection attributes in the query.

Admin → id

Second, the proposed framework hunt in the rules profile down this table. It as of now exists. However this is not the end of the intrusion detection mechanism. The second step is to check for suspicious pivotal magic words in the query. The query as of now contains one of the suspicious pivotal magic words which is single quote.
So XQuery techniques is utilized to fetch questions from the XML document with same from characteristics and same from clause. By contrasting the structure of the query under test and question came back from the XML document we will find that query shouldn't contain the single quote and along these lines it is affirmed as intrusion.

- Select username, password from Admin where id=1 or 1=1- -

The initial phase in the system is to distinguish connection in the middle of tables and selection attributes in the query.

$$Admin \rightarrow id$$
$$Admin \rightarrow 1$$

If we search in our proposed framework we will find that rule for first relation exists but not for the second one therefor this query is affirmed as intrusion.

## VI. CONCLUSION AND FUTURE WORK

Intrusion using web vulnerabilities is a real risk to any organization putting away profitable and classified information in databases. This is progressively all the more so as the quantity of database servers joined with the Internet increments quickly. Existing network-based detection system and also interruption identification frameworks are most certainly not sufficient for recognizing database interruptions. We have presented a framework focused around anomaly and misuse identification for finding SQL injection. We have introduced another encoding strategy for SQL query in XML file as it were empowering the extraction of typical conduct of database application. We then utilized data mining strategy for fingerprinting SQL articulations and use them to recognize SQL Injection intrusions. We want to perform examinations to apply this schema to distinguish its execution in locating assaults and incorporate examinations to different methodologies. This work may be developed to incorporate identification against different assaults like cross site scripting.